\documentclass[a4paper]{article}
\pdfoutput=1 % if your are submitting a pdflatex (i.e. if you have
\usepackage{jcappub} % for details on the use of the package, please
\usepackage[T1]{fontenc} % if needed
\usepackage[utf8]{inputenc}
\usepackage{ulem}
\usepackage{amsmath}
\usepackage{amssymb}
\usepackage{bm}
\usepackage{multirow}
\usepackage{float}
\usepackage{bbold}
\usepackage[compat=1.0.0]{tikz-feynman}

\usepackage{xcolor}
\definecolor{lightpastelpurple}{rgb}{0.69, 0.61, 0.85}
\definecolor{darkpastelred}{rgb}{0.76, 0.23, 0.13}
\definecolor{debianred}{rgb}{0.84, 0.04, 0.33}
\definecolor{palered-violet}{rgb}{0.86, 0.44, 0.58}

\usepackage{graphicx}
\usepackage{hyperref}
\hypersetup{colorlinks=true, linkcolor=teal, urlcolor=blue, citecolor=red}
\usepackage{textcomp}
\usepackage[american]{babel}
\usepackage{enumerate}
\usepackage{xcolor,colortbl}
\usepackage{mathtools}
\usepackage{subcaption}
\usepackage{soul}
\usepackage{tabularx}
\setstcolor{red}
\usepackage{makebox}
\newcommand{\abs}[1]{\left| #1 \right| }

\usepackage{pifont}% http://ctan.org/pkg/pifont

\newcommand\varpm{\mathbin{\vcenter{\hbox{%
  \oalign{\hfil$\scriptstyle\hspace{-0.2ex}+\hspace{-0.2ex}$\hfil\cr
          \noalign{\kern-.5ex}
          $\scriptscriptstyle({-})$\cr}%
}}}}

\newcommand\varmp{\mathbin{\vcenter{\hbox{%
  \oalign{\hfil$\scriptstyle\hspace{-0.2ex}-\hspace{-0.2ex}$\hfil\cr
          \noalign{\kern-.5ex}
          $\scriptscriptstyle({+})$\cr}%
}}}}

\usepackage[capitalise]{cleveref}
\crefname{section}{Sec.}{Secs.}
\crefname{table}{Tab.}{Tabs.}
\crefname{figure}{Fig.}{Figs.}
\crefname{equation}{Eq.}{Eqs.}
\crefname{appendix}{Appendix\ }{Appendix\ }

\graphicspath{{figures/}}

\allowdisplaybreaks

\title{Generating gravitational waveform libraries of exotic compact binaries with deep learning}
\author[a,b]{Felipe F. Freitas}
\author[c,b]{Carlos A.~R.~Herdeiro}
\author[a,b]{Ant\'onio~P.~Morais}
\author[d]{Ant\'onio Onofre}
\author[e]{Roman~Pasechnik}
\author[c,b]{Eugen Radu}
\author[c,b,f]{Nicolas Sanchis-Gual}
\author[g,h]{Rui Santos}

\affiliation[a]{Departamento de F\'{i}sica da Universidade de Aveiro,\\ 
	Campus de Santiago, 3810-183 Aveiro, Portugal.}
\affiliation[b]{Centre  for  Research  and  Development  in  Mathematics  and  Applications  (CIDMA),\\ 
	Campus de Santiago, 3810-183 Aveiro, Portugal.}
\affiliation[c]{Departamento de Matem\'{a}tica da Universidade de Aveiro,\\ 
	Campus de Santiago, 3810-183 Aveiro, Portugal.}
\affiliation[d]{Centro de F\'{i}sica das Universidades do Minho e do Porto (CF-UM-UP),\\
	 Universidade do Minho, 4710-057 Braga, Portugal.}
\affiliation[e]{Department of Astronomy and Theoretical Physics, Lund University,\\ 
	221 00 Lund, Sweden.}
	\affiliation[f]{Departamento de
  Astronom\'{\i}a y Astrof\'{\i}sica, Universitat de Val\`encia,\\
  Dr. Moliner 50, 46100, Burjassot (Val\`encia), Spain}
\affiliation[g]{ISEL -  Instituto Superior de Engenharia de Lisboa,\\
	Instituto Polit\'ecnico de Lisboa  1959-007 Lisboa, Portugal.}
\affiliation[h]{Centro de F\'{i}sica Te\'{o}rica e Computacional, Faculdade de Ci\^{e}ncias,Universidade de Lisboa,\\
	 Campo Grande, Edif\'{i}cio C8 1749-016 Lisboa, Portugal.}

\emailAdd{felipefreitas@ua.pt}
\emailAdd{herdeiro@ua.pt}
\emailAdd{aapmorais@ua.pt}
\emailAdd{Antonio.Onofre@cern.ch}
\emailAdd{roman.pasechnik@thep.lu.se}
\emailAdd{eugen.radu@ua.pt}
\emailAdd{nicolas.sanchis@uv.es}
\emailAdd{rasantos@fc.ul.pt}

\abstract{Current gravitational wave (GW) detections rely on the existence of libraries of theoretical waveforms. Consequently, finding new physics with GWs requires libraries of non-standard models, which are computationally demanding. We discuss how  deep learning frameworks can be used to generate new waveforms "learned" from a simulation dataset obtained, say, from numerical relativity simulations. Concretely, we use the WaveGAN architecture of a generative adversarial network (GAN). As a proof of concept we provide this neural network (NN) with a sample of ($>500$) waveforms from the collisions of exotic compact objects (Proca stars), obtained from numerical relativity simulations. Dividing the sample into a training and a validation set, we show that after a sufficiently large number of training epochs the NN can produce from 12\% to 25\% of the synthetic waveforms with an overlapping match of at least 95\% with the ones from the validation set. We also demonstrate that a NN can be used to predict the overlapping match score, with 90\% of accuracy, of new synthetic samples. These are encouraging results for using GANs for data augmentation and interpolation in the context of GWs, to cover the full parameter space of, say, exotic compact binaries, without the need of intensive numerical relativity simulations.}

\begin{document}
	
	\maketitle
	\flushbottom

%%%%%%%%%%%%%%%%%%%%%%%%%%%%
%%%%%%%%%%%%%%%%%%%%%%%%%%%%
\section{Introduction}
%%%%%%%%%%%%%%%%%%%%%%%%%%%%
%%%%%%%%%%%%%%%%%%%%%%%%%%%%

The advent of the gravitational wave (GW) era~\cite{LIGOScientific:2018mvr,LIGOScientific:2020ibl,LIGOScientific:2021djp} opens new possibilities not only for relativistic astrophysics, cosmology and strong gravity, but also for fundamental physics. Alongside with the unveiling of the population of black holes and neutron stars in the Universe, see $e.g.$,~\cite{LIGOScientific:2018jsj}, it is possible that smoking guns about the nature of dark energy, dark matter and even quantum gravity will emerge from this new channel. In fact, particular events have already established concrete illustrations of how dark energy models can be constrained, see $e.g.$~\cite{Creminelli:2017sry,Baker:2017hug,Ezquiaga:2017ekz} and dark matter could be identified, see $e.g.$~\cite{Bustillo:2020syj}.

The interpretation of GW signals relies on matched filtering~\cite{Owen:1998dk}. Therefore, libraries of theoretical templates are mandatory. The construction of such libraries is a non-trivial, time-consuming process. For the vanilla black hole binary problem in (vacuum) general relativity, the construction of full waveforms, including inspiral, merger and ringdown, became under control after the numerical relativity breakthroughs of 2005 - see the review in~\cite{Pretorius:2007nq}. However, a dense scanning of the full parameter space (of the black hole binary problem) only with numerical relativity simulations is computationally impossible. Thus, a community effort drove the scanning of the parameter space using state of the art numerical relativity simulations, patched together with approximation methods for both the inspiral and the ringdown - see $e.g.$~\cite{Ajith:2007kx,Usman:2015kfa,Boyle:2019kee,Varma:2019csw}, building approximants for the theoretical waveforms with generic parameters.

An under-emphasised caveat of current GW interpretations is the degeneracy problem: can non-standard waveforms fit the data better? Non-standard means waveforms from exotic compact objects, which could either be non-Kerr black holes or horizonless compact objects. Moreover, such exotic compact objects could originate either from general relativity with matter sources or from modified gravity. The difficulty in tackling the above question is, however, the almost complete lack of alternative waveform libraries that can be compared with real events to determine whether the vanilla (Kerr black holes or neutron star binaries) waveforms are indeed the ones selected within a larger library, when employing matched filtering and Bayesian analysis. 

At the time of writing, the one non-standard model of compact binaries for which there has been a more consistent and successful effort to produce waveforms is the case of bosonic ($i.e.$ scalar~\cite{Kaup68,Ruffini69,Schunck:2003kk} or vector~\cite{brito2016proca,Herdeiro:2017fhv,Herdeiro:2019mbz}) stars. The dynamical evolution of these models is theoretically and technically under control~\cite{Liebling:2012fv} and presents a variety of motivations:  bosonic stars emerge in sound physical models, can be dynamically robust~\cite{Liebling:2012fv,sanchis2019nonlinear} and have been put forward as "fuzzy" dark matter~\cite{Hui:2016ltb} lumps and black hole imitators, $e.g.$~\cite{Vincent:2015xta,Olivares:2018abq,Herdeiro:2021lwl}. In the context of GWs, several studies of waveforms from collisions and binaries of bosonic stars have been reported, $e.g.$~\cite{bezares2017final,bezares2018gravitational,bezares2022gravitational,sanchis2019head}. As an application to the ongoing detections, the massive GW event GW190521~\cite{LIGOScientific:2020iuh} was shown to fit well a collision of two vector bosonic ($a.k.a.$ as Proca) stars~\cite{Bustillo:2020syj}. This effort relied on scanning a library of 89 Proca star collision waveforms (in the meantime enlarged to nearly 800 waveform) \cite{catalogue2022}. Still, this only scratches the surface of the full parameter space of the model. As such, looking for efficient computational methodologies that can transform a coarse sampling of the parameter space into a dense coverage is of paramount importance. 

The goal of this paper is to start an exploration of  such a methodology using deep learning techniques. Moreover, the method can, in principle, be used for waveforms produced from generic non-standard compact binaries. Thus, the Proca model explicitly discussed herein can be taken both as interesting in its own right, but simultaneously as a proof of concept of the application of the method, illustrating it but not-exhausting it. To be concrete, we shall be making use of Generative Adversarial Networks (GANs) \cite{2014arXiv1406.2661G}, a particular class of deep learning frameworks.

GANs can be described as unsupervised methods for mapping low-dimensional latent vectors to high-dimensional data. In our case, this means mapping known waveforms, corresponding to a prior distribution, $p_{\rm model}$, to a larger space of waveforms, the generated data distribution, $p_{\rm data}$.  In a nutshell, GANs are based on a game-theoretic scenario where we have two networks competing against each other. On the one hand, we have the \textit{generator}, responsible for mapping the low dimensional vector ${\bf z}$ into the high dimensional samples we want to reproduce ${\bf x} = g({\bf z},\theta^{g})$ ($i.e.$ the waveforms in our case). Here, $\theta^{g}$ are the parameters from the generator network to be adjusted during the training phase. Competing against the generator we have, on the other hand, the \textit{discriminator} network, whose sole purpose is to distinguish between samples drawn from the original dataset and samples drawn from the generator. The discriminator provides a probability,  $d({\bf x};\theta^{d})\in [0.,1.]$, of a given sample ${\bf x}$ being real, as opposed to a fake one drawn from the generator model. Here, $\theta^{d}$ are the parameters from the discriminator network to be adjusted during the training phase.

The simplest way to describe the learning process of a GAN is a zero-sum game, in which a function $\mathcal{L}(\theta^{g},\theta^{d})$ determines the payoff of the discriminator. The generator receives $-\mathcal{L}(\theta^{g}, \theta^{d})$ as its own payoff. During the training phase, each player attempts to maximize its own payoff, so that the generator is trained to maximize $\mathcal{L}(\theta^{g},\theta^{d})$, whereas the discriminator is trained to minimize it. 

\noindent The original proposal \cite{2014arXiv1406.2661G} for the function $\mathcal{L}(\theta^{g}, \theta^{d})$ is~:
\begin{equation}
\label{loss_gan}
   \operatorname{arg} \underset{g}{\operatorname{min}} \  \underset{d}{\operatorname{max}} \ \mathcal{L}(\theta^{g}, \theta^{d}) = \mathbb{E}_{{\bf x}\sim p_{\rm data}} [\log d({\bf x})] + \mathbb{E}_{{\bf z}\sim p_{\rm model}} [\log (1 -d(g(z))] \ ,
\end{equation}
where $\mathbb{E}_{{\bf x}\sim p_{\rm data}}$ and $\mathbb{E}_{{\bf z}\sim p_{\rm model}}$ are the expected values for a sample to be drawn from the data and the generator, respectively.

This drives the discriminator to learn to correctly classify samples as real or fake. Meanwhile, the generator attempts to fool the discriminator by producing fake samples with features as close as possible to the features from real samples. At convergence, the generator's samples are indistinguishable from the real ones, and the discriminator outputs a probability of 50 \% for every sample. The discriminator may be discarded or its parameters can be reused for other purposes later on.

GANs have shown great success in generating high-quality synthetic images \cite{DBLP:journals/corr/abs-2107-13190, DBLP:journals/corr/LedigTHCATTWS16, DBLP:journals/corr/ReedAYLSL16} indistinguishable from real images. This has encouraged the use of GANs for synthetic data generation in broader contexts, in particular in high-energy physics, where in some instances the data generation can be a computational intensive task \cite{Paganini:2017dwg, deOliveira:2017pjk, deOliveira:2017rwa, Paganini:2017hrr}. In this regard, while GANs were developed for image generation \cite{2014arXiv1406.2661G}, there have been attempts to adapt this approach for other formats, such as tabular data \cite{DBLP:journals/corr/abs-2107-13190}, time series \cite{2019arXiv190706673W}, video content augmentation \cite{2021arXiv211110916L} and audio synthesis \cite{2016arXiv160903499V, DBLP:journals/corr/MehriKGKJSCB16, DBLP:journals/corr/abs-1802-04208}. 

In this article, we shall examine the potential of GANs to produce a larger waveform catalogue from a limited dataset of the corresponding waveforms. We shall focus on the case of waveforms produced by Proca star binaries. For this purpose, we shall modify \textit{WaveGAN} \cite{DBLP:journals/corr/abs-1802-04208}, a GAN initially designed to provide an unsupervised synthesis of raw-waveform audio, such that it could learn and produce Proca waveforms from an initial dataset obtained from numerical relativity simulations. Dividing the sample into a training and a validation set, we show that after a sufficiently large number of training epochs the neural network (NN) can generate synthetic data with at least 95\% overlapping match with reference samples from the validation set.

This article is organized as follows. In Sec.~\ref{sec:dataset}, we briefly review the Proca model of bosonic stars and describe the dataset and methodology explored in this study. We also discuss the issue of waveform normalization. In Sec.~\ref{sec:model_arch}, we describe the WaveGAN architecture and the training methodology. Then, in Sec.~\ref{sec:evaluation} we discuss the evaluation methodology, $i.e.$ how to assess the quality of the generated waveforms and use the trained discriminator architectures to predict the match score for new synthetic samples. Sec.~\ref{sec:results} presents our results after applying the chosen architecture, training and evaluation to the initial dataset. Finally, Sec.~\ref{sec:conclusions} provides a final discussion on the approach proposed herein. 

%%%%%%%%%%%%%%%%%%%%%%%%%%%%%%%%%%%%%%%%%%%%%%
%%%%%%%%%%%%%%%%%%%%%%%%%%%%%%%%%%%%%%%%%%%%%%
\section{The Proca model and the  dataset}\label{sec:dataset}
%%%%%%%%%%%%%%%%%%%%%%%%%%%%%%%%%%%%%%%%%%%%%%
%%%%%%%%%%%%%%%%%%%%%%%%%%%%%%%%%%%%%%%%%%%%%%

The Proca stars, their dynamics and the corresponding GWs will be considered in the simplest model: a complex, massive Proca field minimally coupled to Einstein's gravity. The action reads (with $c=1=G$)
\begin{eqnarray}
\label{action}
\mathcal{S}=\int d^4 x \sqrt{-g}
\left [
\frac{R}{16 \pi}
-\frac{1}{4}\mathcal{F}_{\alpha\beta}\bar{\mathcal{F}}^{\alpha\beta}
-\frac{\mu^2}{2}\mathcal{A}_\alpha\bar{\mathcal{A}}^\alpha
\right] \ ,
\end{eqnarray}
where $R$  is the Ricci scalar of the spacetime metric $g$, 
$\mathcal{A}$ is a complex 4-potential, with the field strength $\mathcal{F}_{\alpha\beta} =\partial_\alpha\mathcal{A}_\beta-\partial_\beta\mathcal{A}_\alpha$,
$\mu>0$ corresponds to the mass of the Proca field, and
the overbar denotes complex conjugation.

Spinning Proca stars (the fundamental solutions, in the stable branch~\cite{sanchis2019nonlinear}) can be labelled by their ADM mass, $M\mu$ or, alternatively, by their oscillating frequency $\omega/\mu$, both in units of the Proca field mass. In the following, for simplicity, we shall set $\mu=1$ and label the solutions via $M$. The fundamental solutions in the stable branch have $M$ and $\omega$ in the interval(s)~\cite{Herdeiro:2019mbz}:
\begin{equation}
    (M,\omega)\in ([0,1.125],[0.469,1]) \ .
\end{equation}
Note that the upper (lower) limit in the $M$ interval corresponds to the lower (upper) limit in the $\omega$ interval. The angular momentum of the solutions is determined by $M$. For the considered solutions the total angular momentum is in the range~\cite{Herdeiro:2019mbz} $J\in [0,1.259]$

The collision of two Proca stars generates GWs. These are extracted via the Newman-Penrose (complex) scalar $\Psi_4$. Both the real ($\mathcal{R}(\Psi_4)$) and imaginary parts of this scalar (corresponding to the two GW polarizations) can be decomposed into harmonics. The dominant GW modes, $i.e.$ with higher amplitude, have harmonic indices $(l,m)=(2,2)$ and $(l,m)=(2,0)$. For simplicity, we shall consider only the $(l,m)=(2,2)$ waveforms for each collision, focusing on the real part of the scalar (the "+" polarization). Each waveform is a time series for $r\mathcal{R}(\Psi_4)$, since $\Psi_4$ falls as $1/r$, with $r$ being the distance to the source. 

Our dataset consists of waveforms generated from the merger of two spinning Proca stars with aligned spin axes. This sort of collisions were recently studied in~\cite{Bustillo:2020syj,Sanchis-Gual:2020mzb}. Although the stars start from rest, due to frame dragging the binary describes an eccentric (rather than precisely head-on) trajectory. The end point depends on the progenitor Proca stars. In the region of the parameter space explored here, the Proca star progenitors are sufficiently massive to trigger black hole formation after the merger. 

The waveforms are generated from  numerical evolutions using the \textsc{Einstein toolkit} infrastructure~\cite{EinsteinToolkit,Loffler:2011ay, Zilhao:2013hia}, together with the \textsc{carpet} package \cite{Schnetter:2003rb,Cactus} for mesh-refinement. The Proca evolution equations are solved via a modified Proca thorn \cite{Canuda_2020_3565475, Zilhao:2015tya, sanchis2019head, sanchis2019nonlinear} to include a complex field. We have performed numerical simulations of equal and unequal mass Proca stars. The initial data consists in the superposition of two equilibrium solutions separated by $D = 40/\mu$ \cite{Bustillo:2020syj}, in geometrized units. This guarantees an admissible initial constraint violation. The equilibrium spinning Proca stars are numerically constructed using the solver \textsc{fidisol/cadsol} for non-linear Partial Differential Equations of elliptic type, via a Newton Raphson method - see~\cite{brito2016proca,Herdeiro:2017fhv,Herdeiro:2019mbz} for more details. 

\begin{figure}[H]
    \centering
    \includegraphics[width=\textwidth]{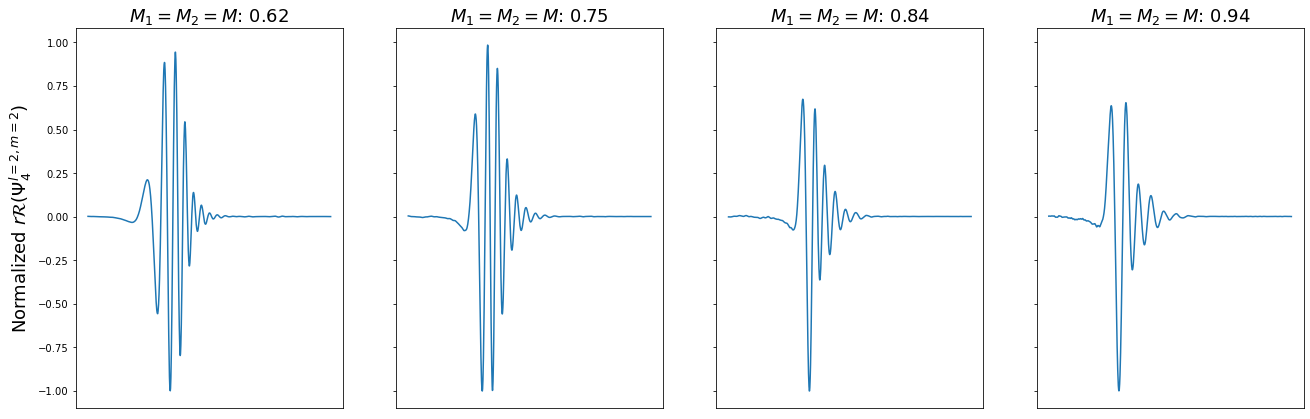}
    \caption{Samples of $r\mathcal{R}(\Psi_{4}^{l=2,m=2})$ for the equal mass data set ($M=M_1=M_2$). The amplitude of $r\mathcal{R}(\Psi_{4}^{l=2,m=2})$ is normalized to be within range of $[-1.,1.]$.}
    \label{fig:sample_equal_masses}
\end{figure}

\begin{figure}[H]
    \centering
    \includegraphics[width=\textwidth]{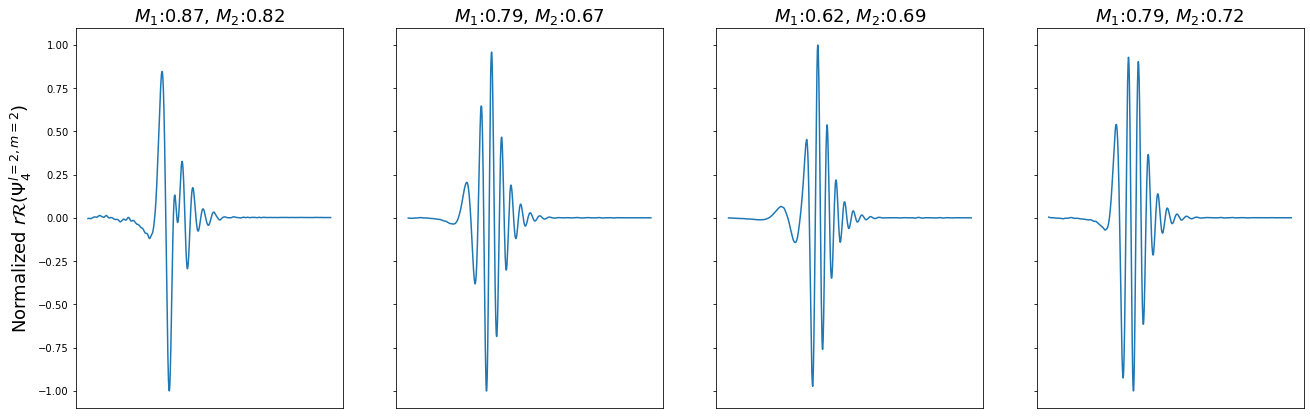}
    \caption{Samples of $r\mathcal{R}(\Psi_{4}^{l=2,m=2})$ for the unequal mass data set ($M_1 \neq M_2$). The amplitude of $r\mathcal{R}(\Psi_{4}^{l=2,m=2})$ is normalized to be within range of $[-1.,1.]$.}
    \label{fig:sample_unequal_masses}
\end{figure}

\begin{figure}[H]
\centering
\begin{subfigure}{.5\textwidth}
  \centering
  \includegraphics[width=\linewidth]{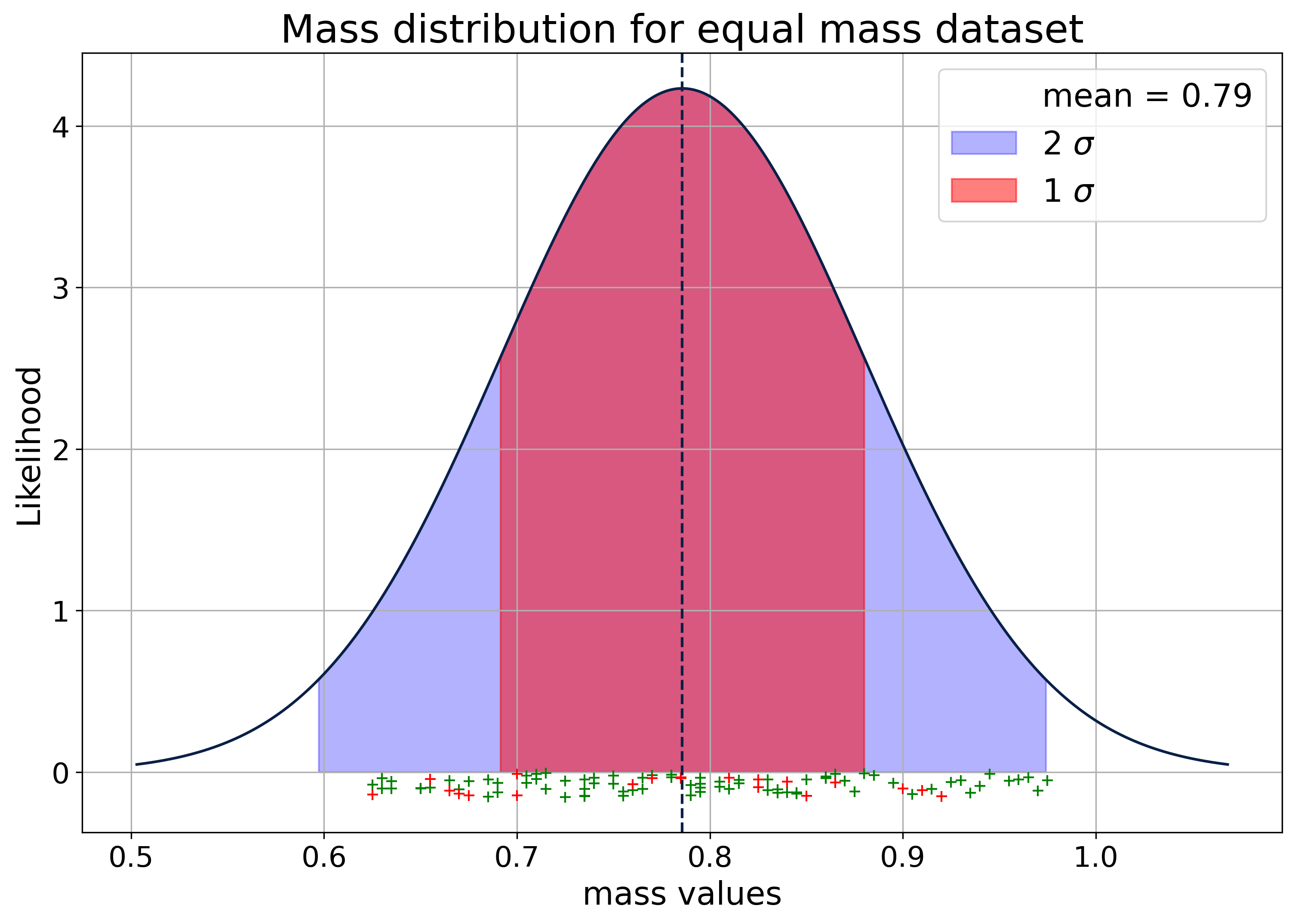}
  \caption{$M_1 = M_2$ dataset. The green (red) crosses represent the samples for the training (validation) set.}
  \label{fig:mass_distro}
\end{subfigure}%
\begin{subfigure}{.5\textwidth}
  \centering
  \includegraphics[width=\linewidth]{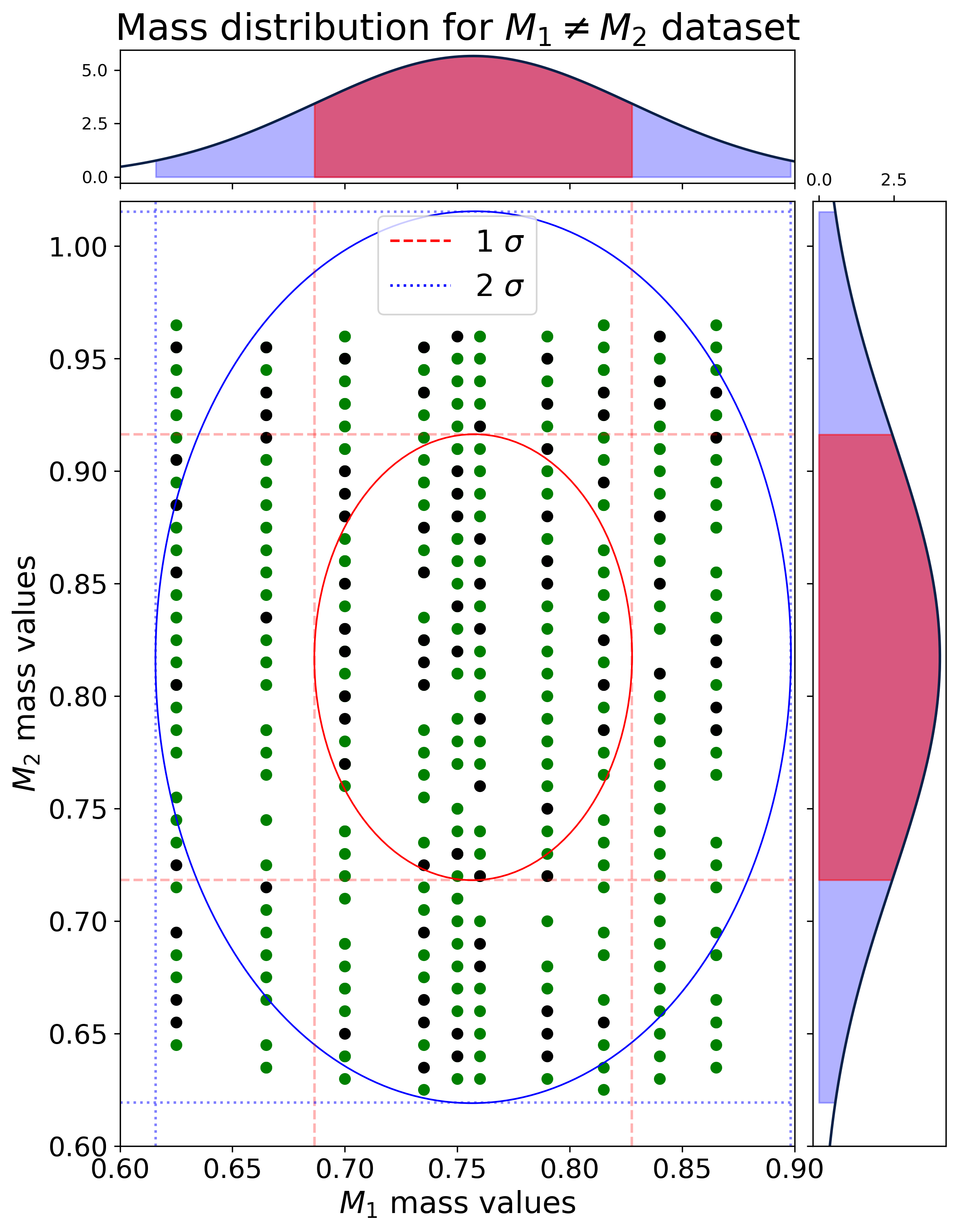}
  \caption{$M_1 \neq M_2$ dataset. The green (black) points represent the samples for the training (validation) set.}
  \label{fig:mass_distro2}
\end{subfigure}
\caption{Mass distribution for the equal (a) and unequal (b) mass dataset with one (red) and two (blue) sigmas. (a) The green and red crosses at the bottom show the sample values for the training and validation datasets, respectively. (b) The green and black dots display the sample values for the training and validation datasets, respectively, while the red and blue ellipsis display the one and two sigmas regions for $M_1$ and $M_2$ distributions. }
\label{fig:mass_distro12}
\end{figure}

We divide our data into two sets: 
\begin{description}
\item[$i)$] one set contains 98 waveforms generated from the merger of two equal mass Proca stars ($M_1 = M_2$). For each collision, we consider waveforms of the $(l,m)=(2,2)$ mode. 
\item[$ii)$] The other dataset consists of 457 waveforms from the merger of two unequal mass Proca stars ($M_1 \neq M_2$), with the same $(l,m)=(2,2)$ mode.
\end{description}
In figure~\ref{fig:sample_equal_masses} and \ref{fig:sample_unequal_masses} some samples for both data sets are illustrated, for the dominant quadrupolar mode $(l,m)=(2,2)$. 
In figure~\ref{fig:mass_distro12} we display the mass distribution for both datasets (equal mass case in figure~\ref{fig:mass_distro} and different mass in figure~\ref{fig:mass_distro2}). Both datasets are pre-processed to be sampled at 2048 Hz. Due to the feature that the dataset have different $y$-ranges we need to normalize them. Having the samples scaled to a similar range helps to prevent or at least mitigate bias and to speed up the optimization process by preventing the model parameter weights to either vanish or explode \cite{279181}. We have tested different methods of scaling, including standard scaling\footnote{$x' = \frac{x - \mu}{\sigma}$, where $x'$ is the transformed feature, $\mu$ and $\sigma$ are the features' average and standard deviation from the dataset, respectively.}, Robust Scaler\footnote{$x' = \frac{x - \mu_{1/2}}{p75 - p25}$, where $x'$ is the transformed feature, $\mu_{1/2}$ is the median and $(p75 - p25)$ is the interquartile range (IQR) which is the difference between the $75^{\rm th}$ and $25^{\rm th}$ percentiles.}, the min-max scaler\footnote{$x'[a,b] = a + \frac{(x - \min{x})(b-a)}{\max{x} - \min{x}}$, where $x'$ is the transformed feature, $a$ and $b$ determine an arbitrary desired interval for the final feature range.} and max-absolute scaler\footnote{$x' = \frac{x}{\max\abs{x}}$.} which scale the features into the $[-1., 1.]$ range without breaking the sparsity of the dataset. We have chosen to scale the datasets according to the max-absolute, since we want to preserve the sparsity of our dataset. It is important to mention that all features are scaled only after the train/validation split happens, to avoid any bias in the training procedure, and their true amplitude range are stored for a later use to transform back the normalized samples into their original values. Then each dataset is shuffled and split into training (80 \% of the total dataset) and validation (20 \% of the total dataset) datasets; these are then fed into the NN model, as further explained in the next Section. 

Each sample consist of a time series representing the real part of the Newman-Penrose scalar $\Psi_{4}^{l=2,m=2}$, together with the value of the mass -- as show in figure~\ref{fig:sample_equal_masses} -- and the feature scale.

%%%%%%%%%%%%%%%%%%%%%%%%%%%%%%%%%%%%%%%%%%%%%%%%
%%%%%%%%%%%%%%%%%%%%%%%%%%%%%%%%%%%%%%%%%%%%%%%%
\section{Model architecture and training methodology}\label{sec:model_arch}
%%%%%%%%%%%%%%%%%%%%%%%%%%%%%%%%%%%%%%%%%%%%%%%%
%%%%%%%%%%%%%%%%%%%%%%%%%%%%%%%%%%%%%%%%%%%%%%%%

Inspired by the use GANs for audio generation, we employ WaveGAN \cite{DBLP:journals/corr/abs-1802-04208} to generate new waveforms "learned" from our simulations dataset. Our purpose is to test whether this method is useful for data augmentation and data interpolation to cover the full parameter space without the need of intensive computational simulations.

The WaveGAN architecture is based on deep convolutional GAN (DCGAN) \cite{2015arXiv151106434R} which popularized the usage of GANs for image synthesis. The DCGAN generator uses the \textit{transposed convolution} operation to iteratively upsample low-resolution feature maps into high-resolution images. The WaveGAN uses a modified \textit{transposed convolution} operation to widen its receptive fields\footnote{The receptive field in Convolutional Neural Networks (CNN) is the region of the input space that affects a particular unit of the network.}. We keep the longer one-dimensional filters of 25, as proposed in~\cite{DBLP:journals/corr/abs-1802-04208}, however we set the number of layers to 4 and channels to 1 at first, and upsample by a factor of 4 at each layer. The discriminator network is modified in a similar fashion, using length-25 filters. The output length from the generator, as well the input length to the discriminator, is set to 2048, to be the same length as the waveforms samples.

The usual GAN generate samples similar to the ones learned in the training. However, this approach is not the most practical if one wants to produce synthetic samples from particular classes present in the dataset, $i.e.$ augment classes or generate samples to interpolate missing regions from the dataset. One way to overcome such problem is to condition our generator and discriminator models. To promote a generator and discriminator to its conditional model forms, one must provide additional information about the training samples, which can be any kind of auxiliary information, such as class labels or, in our case, the mass value $M$ for each sample waveform. We conditioned our WaveGAN using the values of the mass $M$ as labels $\textbf{y}$ and the feature scale $\max\abs{{\bf x}_i}$ used to normalize the sample with the intent to restrict the model to generate samples within these parameter constraints. We can perform the conditioning by feeding $\textbf{y}$ and $\max\abs{{\bf x}_i}$ into both the discriminator and the generator. We use a similar approach as in \cite{DBLP:journals/corr/abs-1809-10636}. To include the label $\textbf{y}$ and $\max\abs{{\bf x}_i}$, we scale the feature maps output from each hidden layers based on the conditioning representation; in our case we scale the feature maps by the mass and feature scale values provided by the sample labels. It is important to mention that this approach is applicable in our case due to the low variance of our labels. In order to deal with high variance labels the best approach is to either normalize them or encode it using a linear layer. Meanwhile, the scale factors $\max\abs{{\bf x}_i}$ help us to constrain the amplitude scale of the synthetic samples, in a sense that when we produce the new samples their amplitude values will be within a region allowed by their physical parameters in the dataset. This is required in order to avoid producing samples which are not permitted by physics or artefacts that can be produced by such methods \cite{odena2016deconvolution}.
\begin{table}[t]
\caption{WaveGAN generator architecture.}
\centering
\footnotesize
\begin{tabular}{ l | l | l }
Operation & Kernel Size & Output Shape \\
\hline
Input $\bm{z} \sim \text{Uniform}(-1, 1)$ &  & ($n$, $100$)\\
Dense 1 & ($100$, $256d$) & ($n$, $256d$)\\
Reshape & & ($n$, $16$, $16d$)\\
ReLU & & ($n$, $16$, $16d$)\\
Trans Conv1D (Stride=$4$) & ($25$, $16d$, $8d$) & ($n$, $64$, $8d$)\\
ReLU & & ($n$, $64$, $8d$)\\
Trans Conv1D (Stride=$4$) & ($25$, $8d$, $4d$) & ($n$, $256$, $4d$)\\
ReLU & & ($n$, $256$, $4d$)\\
Trans Conv1D (Stride=$4$) & ($25$, $4d$, $2d$) & ($n$, $1024$, $2d$)\\
ReLU & & ($n$, $1024$, $2d$)\\
Trans Conv1D (Stride=$4$) & ($25$, $2d$, $d$) & ($n$, $4096$, $d$)\\
ReLU & & ($n$, $4096$, $d$)\\
Trans Conv1D (Stride=$4$) & ($25$, $d$, $c$) & ($n$, $2048$, $c$)\\
Tanh & & ($n$, $2048$, $c$)
\end{tabular}
\label{tab:garch}
\end{table}

\begin{table}[t]
\caption{WaveGAN discriminator architecture.}
\centering
\footnotesize
\begin{tabular}{ l | l | l }
Operation & Kernel Size & Output Shape \\
\hline
Input $\bm{x}$ or $G(\bm{z})$ & & ($n$, $2048$, $c$)\\
Conv1D (Stride=$4$) & ($25$, $c$, $d$) & ($n$, $4096$, $d$) \\
LReLU ($\alpha = 0.2$) & & ($n$, $4096$, $d$) \\
Phase Shuffle ($n = 2$) & & ($n$, $4096$, $d$) \\
Conv1D (Stride=$4$) & ($25$, $d$, $2d$) & ($n$, $1024$, $2d$) \\
LReLU ($\alpha = 0.2$) & & ($n$, $1024$, $2d$) \\
Phase Shuffle ($n = 2$) & & ($n$, $1024$, $2d$) \\
Conv1D (Stride=$4$) & ($25$, $2d$, $4d$) & ($n$, $256$, $4d$) \\
LReLU ($\alpha = 0.2$) & & ($n$, $256$, $4d$) \\
Phase Shuffle ($n = 2$) & & ($n$, $256$, $4d$) \\
Conv1D (Stride=$4$) & ($25$, $4d$, $8d$) & ($n$, $64$, $8d$) \\
LReLU ($\alpha = 0.2$) & & ($n$, $64$, $8d$) \\
Phase Shuffle ($n = 2$) & & ($n$, $64$, $8d$) \\
Conv1D (Stride=$4$) & ($25$, $8d$, $16d$) & ($n$, $16$, $16d$) \\
LReLU ($\alpha = 0.2$) & & ($n$, $16$, $16d$) \\
Reshape & & ($n$, $256d$) \\
Dense & ($256d$, $1$) & ($n$, $1$) \\
\end{tabular}
\label{tab:darch}
\end{table}

Our WaveGAN is implemented in PyTorch\cite{NEURIPS2019_9015}, and we train our model for 1050 epochs using WGAN-GP \cite{DBLP:journals/corr/GulrajaniAADC17} strategy, with ADAM \cite{2014arXiv1412.6980K} as an optimizer, for both generator and discriminator, with learning rate of $10^{-4}$ for the generator and $3\times 10^{-4}$ for the discriminator. We train our networks using batches of size 32, while the validation set has batches of size 16, on a single GPU NVIDIA Tesla V100. As a first task, we set the generators and discriminators to one channel in order to generate the synthetic $r\Psi^{l=2,m=2}$ modes for the equal and different masses datasets. The results for equal and unequal mass datasets are presented in Section~\ref{sec:results}.

%%%%%%%%%%%%%%%%%%%%%%%%%%%%%%%%%%%%%%%%%%%%%%%%%%%%%%
%%%%%%%%%%%%%%%%%%%%%%%%%%%%%%%%%%%%%%%%%%%%%%%%%%%%%%
\section{Evaluation methodology}\label{sec:evaluation}
%%%%%%%%%%%%%%%%%%%%%%%%%%%%%%%%%%%%%%%%%%%%%%%%%%%%%%
%%%%%%%%%%%%%%%%%%%%%%%%%%%%%%%%%%%%%%%%%%%%%%%%%%%%%%

\begin{figure}[H]
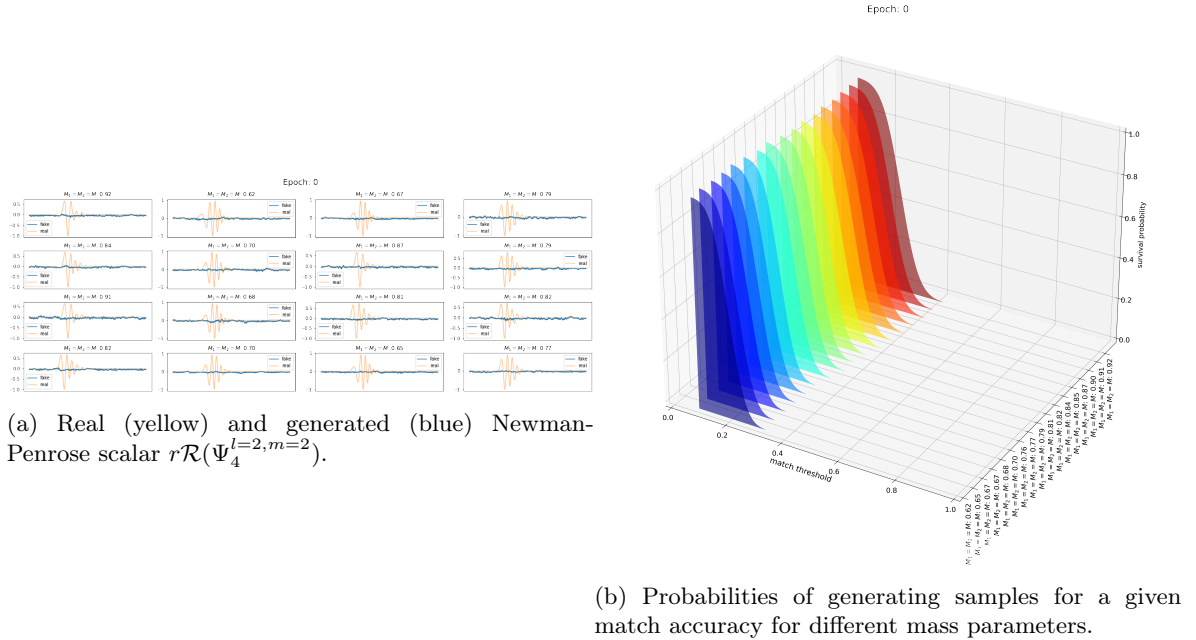

\centering
\begin{subfigure}{.5\textwidth}
  \centering
  \includegraphics[width=\linewidth]{figures/equal mass/train evolution/sample_waves_0.pdf}
  \caption{Real (yellow) and generated (blue) Newman-Penrose scalar $r\mathcal{R}(\Psi_{4}^{l=2,m=2})$.}
  \label{fig:gen_vs_real_epoch_0}
\end{subfigure}%
\begin{subfigure}{.5\textwidth}
  \centering
  \includegraphics[width=0.9\linewidth]{figures/equal mass/train evolution/3d_survival_matches_epoch_0.pdf}
  \caption{Probabilities of generating samples for a given match accuracy for different mass parameters.}
  \label{fig:probs_epoch_0}
\end{subfigure}
\caption{Equal mass dataset, epoch 0. Evaluation of the quality of the generated samples compared to the real samples of  $r\mathcal{R}(\Psi_{4}^{l=2,m=2})$ and probabilities to generate good quality samples.}
\label{fig:epoch_0}
\end{figure}

The evaluation of generative models is an ongoing topic in the community \cite{DBLP:journals/corr/abs-2103-09396}. Just as important as choosing the right strategy to train a generative model, is selecting the right metric to evaluate the quality of the generated samples. 
A direct comparison between the synthetic samples and the real ones can be a useful diagnostic, often allowing us to build intuition of how the generative model is working, how it is failing and how it can be improved. However, qualitative as well quantitative analysis based on this approach can be misleading about the performance of the generator. In order to evaluate the quality, and therefore how  trustworthy is our generator, we shall employ the following strategy. Using the PyCBC \cite{alex_nitz_2020_4075326} matched filtering module, we estimate the overlap over time and phase between the synthetic and real samples for a given value of the  parameters. We generate a set of 1000 synthetic samples for a given set of parameters,  ($M$) for equal mass dataset or ($M_1$, $M_2$) for unequal mass dataset, and compute the overlapping match for each sample to the real equivalent samples. The overlapping match is computed with the real and synthetic normalized samples, so we can ensure that features generated for the synthetic samples are as close as possible to the expected real features. With these values, we estimate the probability of a generated sample to be above a certain threshold of match. In figures~\ref{fig:epoch_0}-\ref{fig:Uneq_epoch_1000} we show the evolution of the match between synthetic and original samples throughout the training epochs of our NN model. To visualize the overlapping between the real and synthetic samples, we plot various samples from the equal and different mass datasets and select synthetic samples according to their overlapping matches to check against the real ones; these plots are shown in figure~\ref{fig:samples_synth_equal_mass} and figure~\ref{fig:samples_synth_unequal_mass}.

Using the overlapping match we build a separate dataset with synthetic samples and their respective match score. This dataset is further used to train another NN with the intent of predict the match score for a given waveform sample. This new dataset consists of 85000 synthetic samples, and their matched scores are evaluated using the validation dataset for the equal and unequal mass dataset, each sample consisting of the normalized $r\Psi_{4}^{l=2,m=2}$ time-series, the value of the match score for the synthetic sample when compared to the real one and the mass $M$ parameter value. This dataset is again randomly shuffled and divided by 80\% for training and 20\% for validation dataset.

\begin{figure}[h]
\centering
\includegraphics[width=\linewidth]{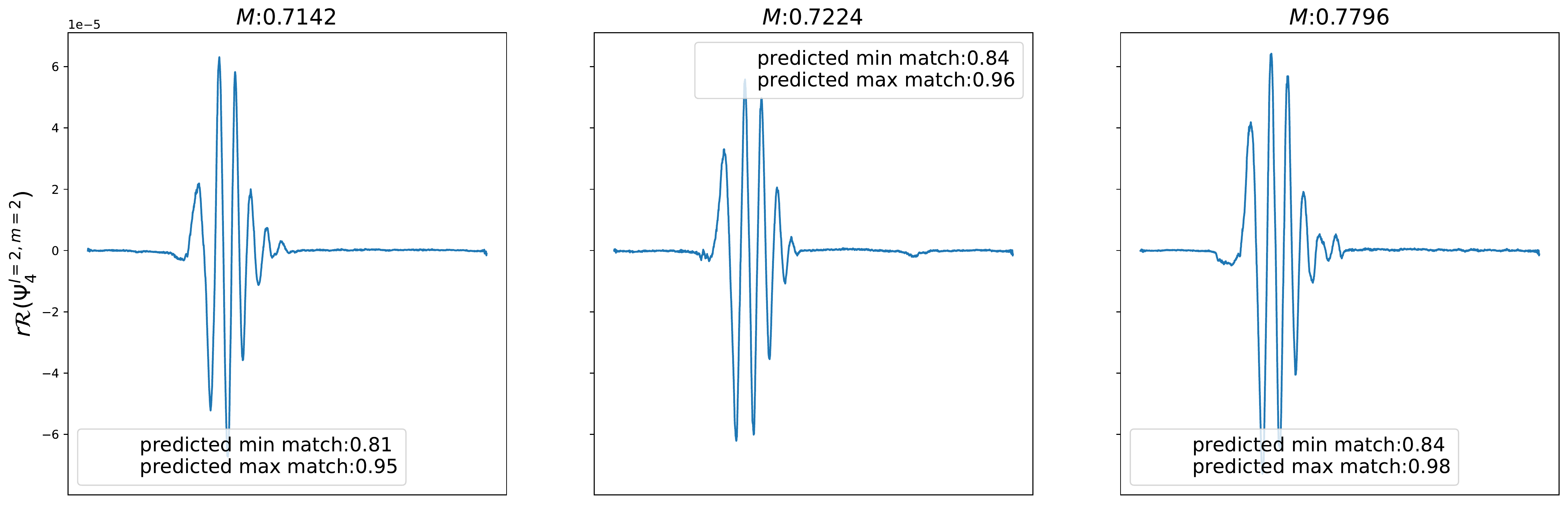}
\caption{ Synthetic samples and their predicted matched samples for parameter values of $M$ not present, but within the one sigma range from the mean, in the current equal mass dataset.}
\label{fig:synth_sample_1}
\end{figure}

\begin{figure}[h]
\centering
\includegraphics[width=\linewidth]{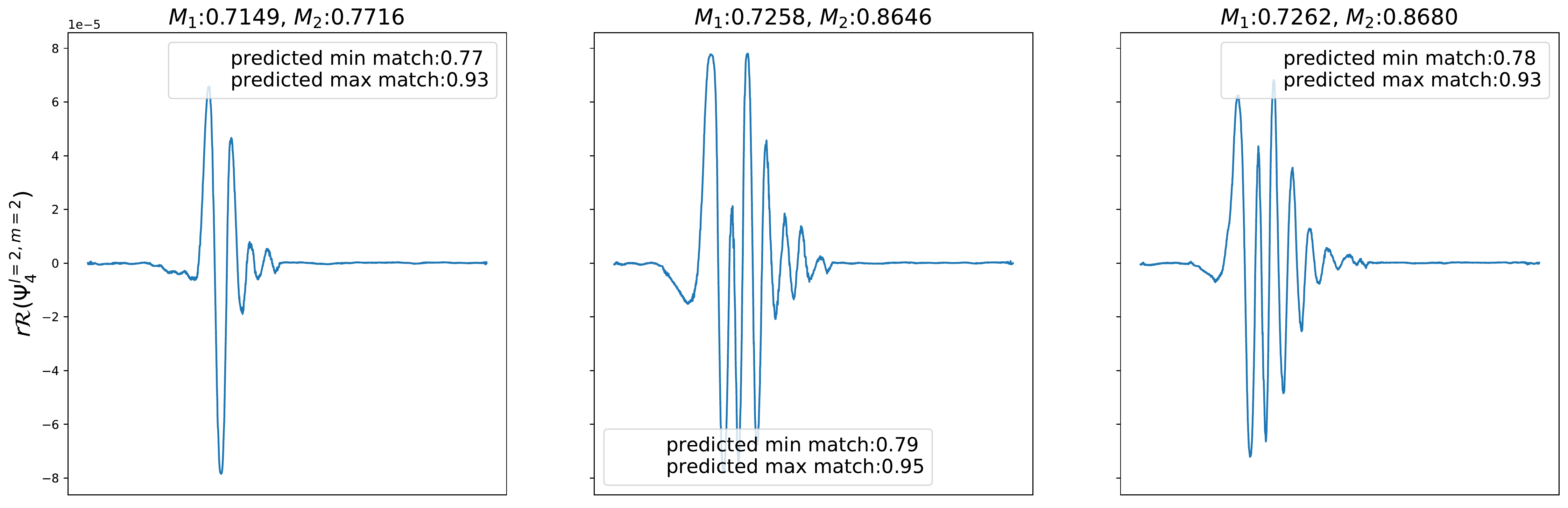}
\caption{ Synthetic samples and their predicted matched samples for parameter values of $M_1$ and $M_2$ not present, but within the one sigma range from the mean, in the current unequal mass dataset.}
\label{fig:synth_sample_2}
\end{figure}

We use the discriminator architecture as in Table \ref{tab:darch}, with the inclusion of dropout layers, with 0.2 probability of an element to be zeroed, in between each convolution layer, the dropout layers are included so we can further use the Monte-Carlo dropout method \cite{2017arXiv170807120S} to estimate the predictions' uncertainties. We train this modified discriminator using the mean squared loss since the new task now - to predict the match score for a given sample - is similar to a regression problem. We also take advantage of the transfer learning method and use the weights from the trained discriminators to speed up the training of the new NN. We use the Adam optimizer and this time we train the model using the OneCycle \cite{2017arXiv170807120S} learning rate policy; the model is trained for 11 epochs with an initial learning rate $2e^{-3}$ and uses the root mean squared error as main metric. The final model can predict the match values for a sample with an accuracy around 90\%. Using the Monte-Carlo dropout scheme we can further estimate the uncertainty of the model and determine the minimum and maximum predicted values of matches. Figures~\ref{fig:synth_sample_1} and \ref{fig:synth_sample_2} display three synthetic samples generated from our model using $M$ values which are not present in the original dataset, but within the one sigma range from the mean $M$ value, and their minimum and maximum predicted match values.

%%%%%%%%%%%%%%%%%%%%%%%%%%%%%%%%%%%%%%%%%%%%%%%%%%%%%%
%%%%%%%%%%%%%%%%%%%%%%%%%%%%%%%%%%%%%%%%%%%%%%%%%%%%%%
\section{Results and discussions}\label{sec:results}
%%%%%%%%%%%%%%%%%%%%%%%%%%%%%%%%%%%%%%%%%%%%%%%%%%%%%%
%%%%%%%%%%%%%%%%%%%%%%%%%%%%%%%%%%%%%%%%%%%%%%%%%%%%%%

\begin{figure}[H]
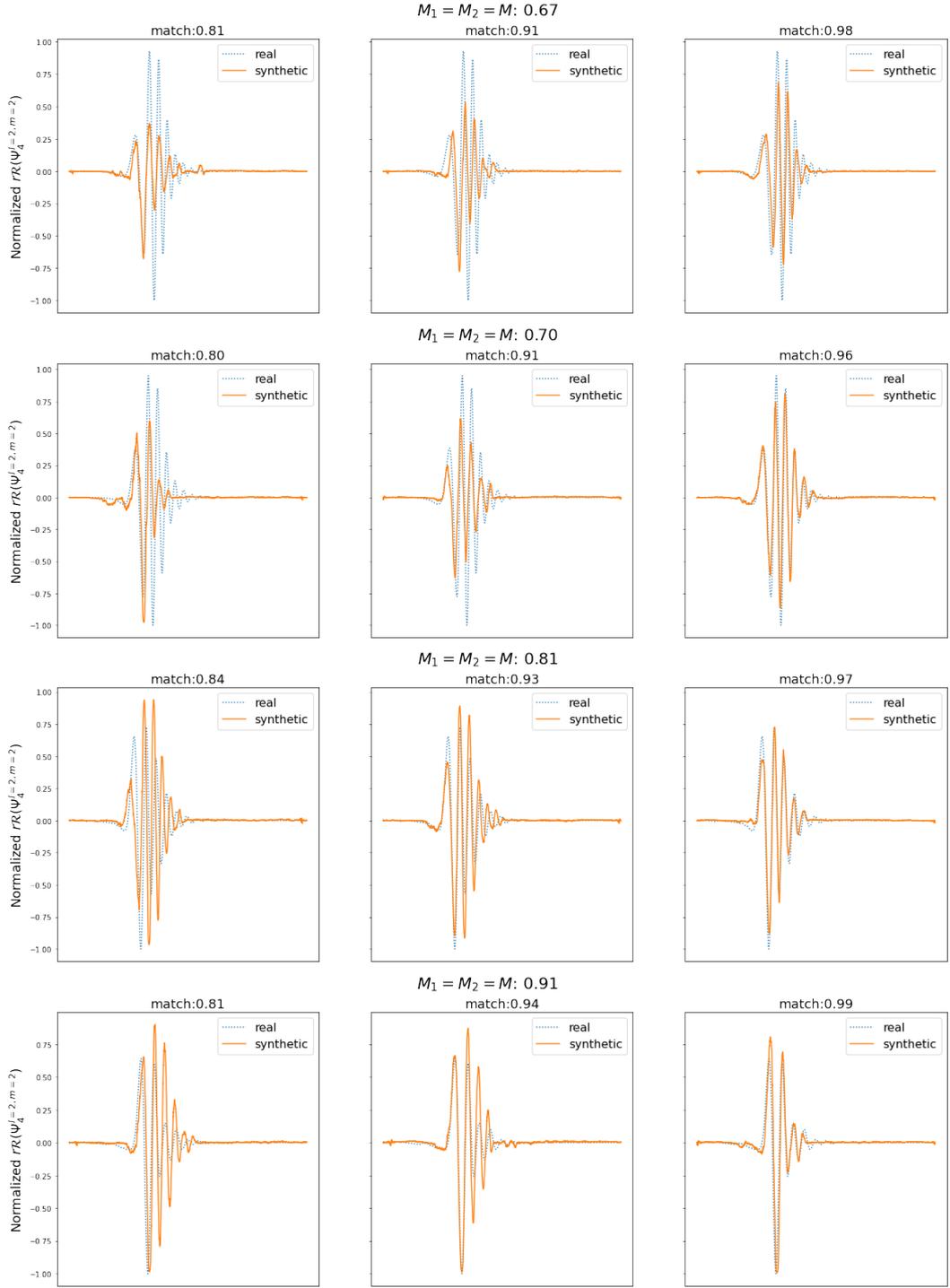

\centering
\includegraphics[width=0.9\linewidth]{figures/equal mass/real and fake/fake_and_real_m_0d67_0d9200.pdf} \\
\includegraphics[width=0.9\linewidth]{figures/equal mass/real and fake/fake_and_real_m_0d70_0d9100.pdf} \\
\includegraphics[width=0.9\linewidth]{figures/equal mass/real and fake/fake_and_real_m_0d81_0d8725.pdf} \\
\includegraphics[width=0.9\linewidth]{figures/equal mass/real and fake/fake_and_real_m_0d91_0d8300.pdf} 
\caption{Real (dotted blue) and generated (continuous yellow)  $r\mathcal{R}(\Psi_{4}^{l=2,m=2})$ for overlapping matches in the regions $0.8 \leq \text{match} \leq 0.85$ (left column), $0.9 \leq \text{match} \leq 0.95$ (middle column) and $\text{match} > 0.95$ (right column) of the equal mass dataset.}
\label{fig:samples_synth_equal_mass}
\end{figure}

\begin{figure}[H]
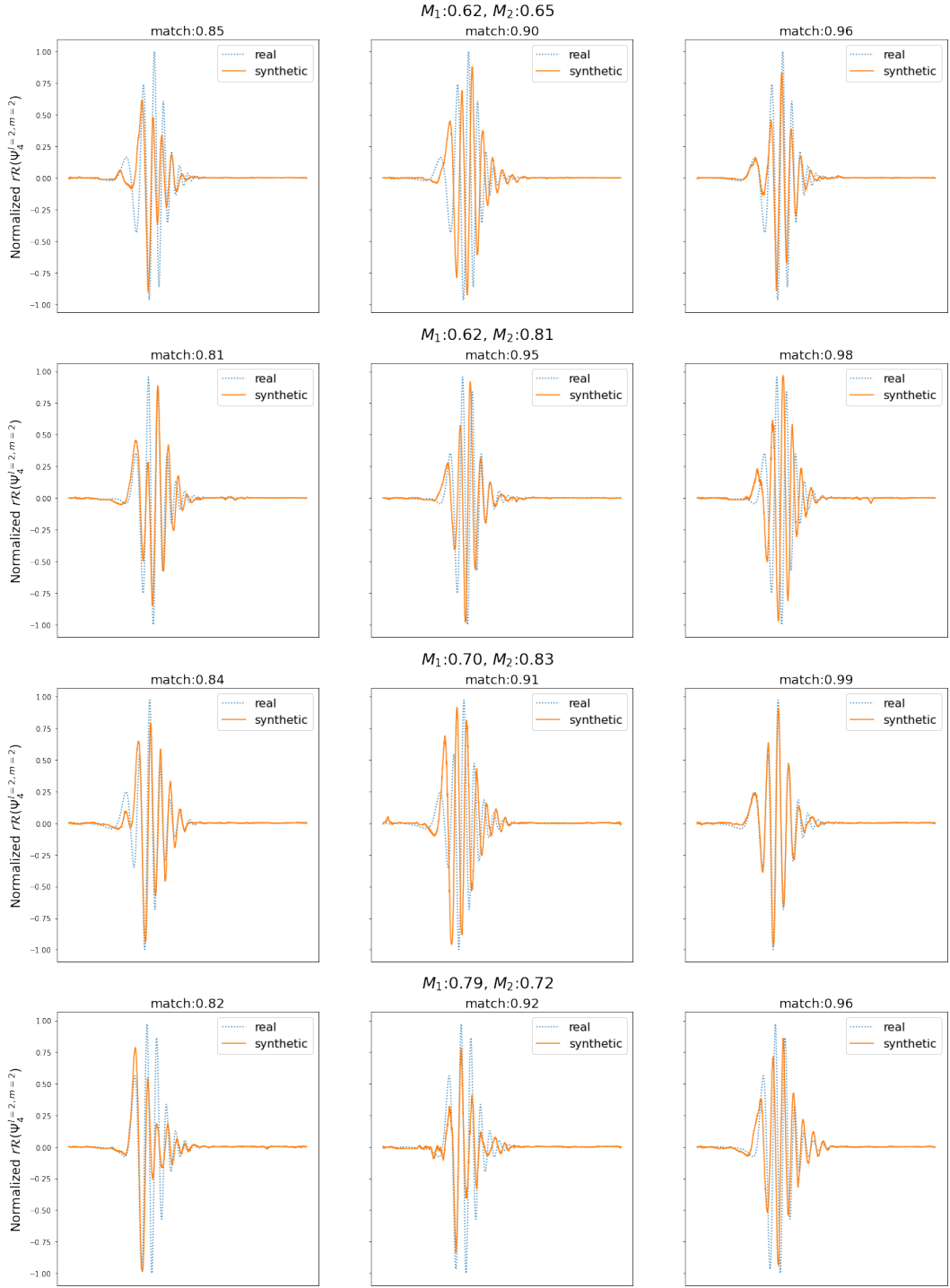

\centering
\includegraphics[width=0.9\linewidth]{figures/unequal mass/real and fake/fake_and_real_m_0d62_0d65_0d925.pdf} \\
\includegraphics[width=0.9\linewidth]{figures/unequal mass/real and fake/fake_and_real_m_0d62_0d81_0d875.pdf} \\
\includegraphics[width=0.9\linewidth]{figures/unequal mass/real and fake/fake_and_real_m_0d70_0d83_0d868.pdf} \\
\includegraphics[width=0.9\linewidth]{figures/unequal mass/real and fake/fake_and_real_m_0d79_0d72_0d908.pdf} 
\caption{Same as figure~\ref{fig:samples_synth_unequal_mass}
but for the unequal mass dataset.}
\label{fig:samples_synth_unequal_mass}
\end{figure}

The results of our evaluation are shown in Tables~\ref{tab:equal_mass_quality} and \ref{tab:unequal_mass_2}. In figures~\ref{fig:samples_synth_equal_mass} and \ref{fig:samples_synth_unequal_mass} we sample waveforms and rank them according to their respective matches to their waveform reference, from the validation dataset. We are able to generate waveforms with a 95\% to 99\% match with the reference waveforms, despite the model never having had access to the samples from the validation dataset.
From Tables~\ref{tab:equal_mass_quality} and \ref{tab:unequal_mass_2}, we see an interesting and understandable pattern: our model has a higher chance of producing samples which are closer to the expected ones whenever the sample parameters desired are close to the mean of the mass $M$ parameters of the dataset, $i.e.$ the region where we have more samples and, by and large, the model is "learning" the underlying features which appear more often in the data. Nevertheless, in the parameter region away from the expected mean we still have a relatively high chance of producing samples with, at least, 90\% match with the expected real samples. Such results show the potential of the method presented in this work. Moreover, one key aspect of such methodology that should be emphasised is the speed of generation of the samples. By using this technique we can generate 1000 synthetic samples in 188 ms\footnote{Benchmark obtained in a Intel\textregistered Core$^{\text{TM}}$ i7-8750H CPU with an NVIDIA GeForce GTX 1070.}, assuming that we can have from 16\% to 25\% (depending on the mass parameter values) samples which are 95\% similar to the expected real waveforms. In other words, we can quickly build a catalogue of waveforms, to bridge the gaps within the parameter space. Such numbers show a clear advantage of the method to help the task of exploring the parameter space for the case of Proca stars waveforms, or more generically, gravitational waveforms generation in any model.

Although the methodology put forward here shows promising results, one should consider this proposal as a proof of concept, with necessary improvements yet to be done. The evaluation methodology using a NN can be improved by experimenting with different architectures for this task, and use Bayesian methods to estimate the error on the predicted match score. Additionally, the number of samples is a factor that heavily influences the quality of the synthetic sample; a rich dataset with not only more samples but samples with different sample rates can greatly improve the quality of the synthetic samples. New architecture models are already being tested and the Transformer~\cite{2017arXiv170603762V} based architectures are showing promising results that will be explored in a future project.

\begin{table}[h!]
\centering
\begin{tabular}{ |c|c|c|c| } 
 \hline
 $(M_1 = M_2 = M)$ & $P(0.8)$ & $P(0.9)$ & $P(0.95)$ \\
 \hline\hline

    \cellcolor{lightpastelpurple} 0.62 & 0.82 & 0.38 & 0.16 \\
    \hline
    \cellcolor{lightpastelpurple} 0.65 & 0.67 & 0.28 & 0.12 \\
    \hline
    \cellcolor{lightpastelpurple} 0.67 & 0.70 & 0.31 & 0.13 \\
    \hline
    \cellcolor{lightpastelpurple} 0.67 & 0.67 & 0.29 & 0.12 \\
    \hline
    \cellcolor{palered-violet} 0.68 & 0.66 & 0.28 & 0.11 \\
    \hline
    \cellcolor{palered-violet} 0.70 & 0.70 & 0.32 & 0.13 \\
    \hline
    \cellcolor{palered-violet} 0.76 & 0.89 & 0.47 & 0.20 \\
    \hline
    \cellcolor{palered-violet} 0.77 & 0.93 & 0.52 & 0.22 \\
    \hline
    \cellcolor{palered-violet} 0.79 & 0.96 & 0.57 & 0.24 \\
    \hline
    \cellcolor{palered-violet} 0.81 & 0.97 & 0.60 & 0.24 \\
    \hline
    \cellcolor{palered-violet} 0.82 & 0.97 & 0.55 & 0.21 \\
    \hline
    \cellcolor{palered-violet} 0.84 & 0.97 & 0.54 & 0.20 \\
    \hline
    \cellcolor{palered-violet} 0.85 & 0.90 & 0.44 & 0.17 \\
    \hline
    \cellcolor{lightpastelpurple} 0.87 & 0.79 & 0.34 & 0.13 \\
    \hline
    \cellcolor{lightpastelpurple} 0.90 & 0.61 & 0.23 & 0.09 \\
    \hline
    \cellcolor{lightpastelpurple} 0.91 & 0.59 & 0.22 & 0.09 \\
    \hline
    \cellcolor{lightpastelpurple} 0.92 & 0.52 & 0.19 & 0.07 \\
    \hline
\end{tabular}
\caption{Probabilities for a generator to produce  $r\mathcal{R}(\Psi_{4}^{l=2,m=2})$ with an overlapping match to the original one above 0.8 (second column), 0.9 (third column) and 0.95 (last column) for each mass $(M = M_1 = M_2)$ and $\omega$ values of the validation equal mass dataset. The colors show where the values fall into the 1 and 2 $\sigma$ mass distribution in figure~\ref{fig:mass_distro}.}
\label{tab:equal_mass_quality}
\end{table}

\begin{table}[h!]
\centering
\begin{tabular}{ |c|c|c|c|c| } 
 \hline
 $M_1$ & $M_2$ & $P(0.8)$ & $P(0.9)$ & $P(0.95)$ \\
 \hline\hline
\cellcolor{lightpastelpurple} 0.62 & \cellcolor{lightpastelpurple} 0.65 & 0.57 & 0.24 & 0.10 \\
\hline
\cellcolor{lightpastelpurple} 0.62 & \cellcolor{lightpastelpurple} 0.69 & 0.59 & 0.24 & 0.10 \\
\hline
\cellcolor{lightpastelpurple} 0.62 & \cellcolor{palered-violet} 0.81 & 0.57 & 0.25 & 0.11 \\ 
\hline
\cellcolor{lightpastelpurple} 0.62 & \cellcolor{lightpastelpurple} 0.95 & 0.69 & 0.32 & 0.14 \\
\hline
\cellcolor{palered-violet} 0.70 & \cellcolor{palered-violet} 0.83 & 0.50 & 0.21 & 0.09 \\ 
\hline
\cellcolor{palered-violet} 0.70 & \cellcolor{palered-violet} 0.90 & 0.60 & 0.19 & 0.06 \\ 
\hline
\cellcolor{palered-violet} 0.74 & \cellcolor{lightpastelpurple} 0.67 & 0.49 & 0.22 & 0.10 \\ 
\hline
\cellcolor{palered-violet} 0.74 & \cellcolor{lightpastelpurple} 0.69 & 0.46 & 0.20 & 0.09 \\ 
\hline
\cellcolor{palered-violet} 0.74 & \cellcolor{palered-violet} 0.88 & 0.83 & 0.32 & 0.11 \\ 
\hline
\cellcolor{palered-violet} 0.76 & \cellcolor{lightpastelpurple} 0.68 & 0.62 & 0.28 & 0.12 \\ 
\hline
\cellcolor{palered-violet} 0.76 & \cellcolor{lightpastelpurple} 0.69 & 0.53 & 0.23 & 0.10 \\ 
\hline
\cellcolor{palered-violet} 0.79 & \cellcolor{lightpastelpurple} 0.64 & 0.46 & 0.18 & 0.07 \\ 
\hline
\cellcolor{palered-violet} 0.79 & \cellcolor{palered-violet} 0.72 & 0.70 & 0.33 & 0.15 \\ 
\hline
\cellcolor{palered-violet} 0.79 & \cellcolor{lightpastelpurple} 0.93 & 0.66 & 0.22 & 0.08 \\ 
\hline
\cellcolor{palered-violet} 0.81 & \cellcolor{lightpastelpurple} 0.65 & 0.52 & 0.24 & 0.11 \\ 
\hline
\cellcolor{palered-violet} 0.81 & \cellcolor{palered-violet} 0.79 & 0.61 & 0.25 & 0.10 \\ 
\hline
\cellcolor{lightpastelpurple} 0.87 & \cellcolor{palered-violet} 0.80 & 0.84 & 0.35 & 0.12 \\ 
\hline
\cellcolor{lightpastelpurple} 0.87 & \cellcolor{palered-violet} 0.81 & 0.81 & 0.30 & 0.10 \\ 
\hline
\cellcolor{lightpastelpurple} 0.87 & \cellcolor{lightpastelpurple} 0.94 & 0.37 & 0.11 & 0.04 \\
\hline
\cellcolor{lightpastelpurple} 0.87 & \cellcolor{lightpastelpurple} 0.94 & 0.43 & 0.13 & 0.04 \\
\hline
\end{tabular}
\caption{Same as Table~\ref{tab:equal_mass_quality} but for the unequal mass case providing
each mass $(M_1 \neq M_2)$ and values of the validation dataset. The colors show where the values fall into the 1 and 2 $\sigma$ mass distribution in figure~\ref{fig:mass_distro2}.}
\label{tab:unequal_mass_2}
\end{table}

\begin{figure}[H]
\centering
\begin{subfigure}{.5\textwidth}
  \centering
  \includegraphics[width=\linewidth]{figures/unequal mass/train evolution/sample_waves_unequal_mass_0.pdf}
  \caption{Real (yellow) and generated (blue) Newman-Penrose scalar $\Psi_{4}^{l=2,m=2}$.}
  \label{fig:Uneq_gen_vs_real_epoch_0}
\end{subfigure}%
\begin{subfigure}{.5\textwidth}
  \centering
  \includegraphics[width=0.9\linewidth]{figures/unequal mass/train evolution/3d_survival_matches_epoch_0.pdf}
  \caption{Probabilities of generating samples for a given match accuracy, for different mass parameters.}
  \label{fig:Uneq_probs_epoch_0}
\end{subfigure}
\caption{Evaluation of the quality of the generated samples compared to the real samples of the Newman-Penrose scalar $\Psi_{4}^{l=2,m=2}$ and probabilities to generate good quality samples at epoch 0 with a different mass dataset.
}
\label{fig:Uneq_epoch_0}
\end{figure}

\begin{figure}[H]
\centering
\begin{subfigure}{.5\textwidth}
  \centering
  \includegraphics[width=\linewidth]{figures/equal mass/train evolution/sample_waves_500.pdf}
  \caption{Real (yellow) and generated (blue) Newman-Penrose scalar $\Psi_{4}^{l=2,m=2}$.}
  \label{fig:gen_vs_real_epoch_500}
\end{subfigure}%
\begin{subfigure}{.5\textwidth}
  \centering
  \includegraphics[width=0.9\linewidth]{figures/equal mass/train evolution/3d_survival_matches_epoch_500.pdf}
  \caption{Probabilities of generating samples for a given match accuracy for different mass parameters.}
  \label{fig:probs_epoch_500}
\end{subfigure}
\caption{Evaluation of the quality of the generated samples compared to the real samples of the Newman-Penrose scalar $\Psi_{4}^{l=2,m=2}$ and probabilities to generate good quality samples at epoch 500.
}
\label{fig:epoch_500}
\end{figure}

\begin{figure}[H]
\centering
\begin{subfigure}{.5\textwidth}
  \centering
  \includegraphics[width=\linewidth]{figures/unequal mass/train evolution/sample_waves_unequal_mass_500.pdf}
  \caption{Real (yellow) and generated (blue) Newman-Penrose scalar $\Psi_{4}^{l=2,m=2}$.}
  \label{fig:Uneq_gen_vs_real_epoch_500}
\end{subfigure}%
\begin{subfigure}{.5\textwidth}
  \centering
  \includegraphics[width=0.9\linewidth]{figures/unequal mass/train evolution/3d_survival_matches_epoch_500.pdf}
  \caption{Probabilities of generating samples for a given match accuracy for different mass parameters.}
  \label{fig:Uneq_probs_epoch_500}
\end{subfigure}
\caption{Evaluation of the quality of the generated samples compared to the real samples of the Newman-Penrose scalar $\Psi_{4}^{l=2,m=2}$ and probabilities to generate good quality samples at epoch 500 with the different mass dataset.
}
\label{fig:Uneq_epoch_500}
\end{figure}

\begin{figure}[H]
\centering
\begin{subfigure}{.5\textwidth}
  \centering
  \includegraphics[width=\linewidth]{figures/equal mass/train evolution/sample_waves_1050.pdf}
  \caption{Real (yellow) and generated (blue) Newman-Penrose scalar $\Psi_{4}^{l=2,m=2}$.}
  \label{fig:gen_vs_real_epoch_1050}
\end{subfigure}%
\begin{subfigure}{.5\textwidth}
  \centering
  \includegraphics[width=0.9\linewidth]{figures/equal mass/train evolution/3d_survival_matches_epoch_1050.pdf}
  \caption{Probabilities of generating samples for a given match accuracy for different mass parameters.}
  \label{fig:probs_epoch_1050}
\end{subfigure}
\caption{Evaluation of the quality of the generated samples compared to the real samples of the Newman-Penrose scalar $\Psi_{4}^{l=2,m=2}$ and probabilities to generate good quality samples at epoch 1050.
}
\label{fig:epoch_1050}
\end{figure}

\begin{figure}[H]
\centering
\begin{subfigure}{.5\textwidth}
  \centering
  \includegraphics[width=\linewidth]{figures/unequal mass/train evolution/sample_waves_unequal_mass_1050.pdf}
  \caption{Real (yellow) and generated (blue) Newman-Penrose scalar $\Psi_{4}^{l=2,m=2}$.}
  \label{fig:Uneq_gen_vs_real_epoch_1000}
\end{subfigure}%
\begin{subfigure}{.5\textwidth}
  \centering
  \includegraphics[width=0.9\linewidth]{figures/unequal mass/train evolution/3d_survival_matches_epoch_1050.pdf}
  \caption{Probabilities of generating samples for a given match accuracy for different mass parameters.}
  \label{fig:Uneq_probs_epoch_1000}
\end{subfigure}
\caption{Evaluation of the quality of the generated samples compared to the real samples of the Newman-Penrose scalar $\Psi_{4}^{l=2,m=2}$ and probabilities to generate good quality samples at epoch 1050 with the different mass dataset.
}
\label{fig:Uneq_epoch_1000}
\end{figure}

%%%%%%%%%%%%%%%%%%%%%%%%%%%%%%%%%%%%%%%%%%%%%%%%%%%%%%
%%%%%%%%%%%%%%%%%%%%%%%%%%%%%%%%%%%%%%%%%%%%%%%%%%%%%%
\section{Conclusions}\label{sec:conclusions}
%%%%%%%%%%%%%%%%%%%%%%%%%%%%%%%%%%%%%%%%%%%%%%%%%%%%%%
%%%%%%%%%%%%%%%%%%%%%%%%%%%%%%%%%%%%%%%%%%%%%%%%%%%%%%

In this work, we have presented a unique application of a particular GAN architecture - WaveGan - in the context of unsupervised gravitational waveforms generation. The model presented here can generates hundreds of thousands of waveforms within very short time intervals, labelled by the physical parameters of the Proca stars that sourced them -- with equal ($M$) or different ($M_1, M_2$) masses -- with a high probability that such "fake" samples are 95\%, or higher, similar to the expected real ones.

We have also explored the use of the trained discriminator architectures to assist in the task of estimating the overlap score of synthetic samples, which can be used to select the synthetic waveforms which show closer features to the expected real ones. The methods presented here show that it is possible to use such techniques to accelerate the generation of the waveforms, in particular for the case of binaries of exotic compact objects.

In a future work we plan to extend and refine the method to produce samples with higher quality and automatically assign the overlapping match factor. We are also working on applying such methods in the generation of waveforms from core-collapse supernovae.

\bigskip

%%%%%%%%%%%%%%%%%%%%%%%%%%%
\acknowledgments
%%%%%%%%%%%%%%%%%%%%%%%%%%%

This work is supported by the Center for Research and Development in Mathematics and Applications (CIDMA) through the Portuguese Foundation for Science and Technology (FCT - Funda\c c\~ao para a Ci\^encia e a Tecnologia), references UIDB/04106/2020 and, UIDP/04106/2020, and by national funds (OE), through FCT, I.P., in the scope of the framework contract foreseen in the numbers 4, 5 and 6 of the article 23, of the Decree-Law 57/2016, of August 29, changed by Law 57/2017, of July 19. This work is also supported by CFTC-UL through FCT, references UIDB/00618/2020 and UIDP/00618/2020.
The author(s) gratefully acknowledges the computer resources at Artemisa, funded by the European Union ERDF and Comunitat Valenciana as well as the technical support provided by the Instituto de Física Corpuscular, IFIC (CSIC-UV).
We acknowledge support  from  the  projects PTDC/FIS-OUT/28407/2017, PTDC/FIS-PAR/31000/2017, CERN/FIS-PAR/0027/2019, CERN/FIS-PAR/0002/2019 and PTDC/FIS-AST/3041/2020. This work has further been supported by the European Union’s Horizon 2020 research and innovation (RISE) programme H2020-MSCA-RISE-2017 Grant No. FunFiCO-777740. NSG was also supported by the Spanish Ministerio de Universidades, reference UP2021-044, within the European Union-Next Generation EU. 
This work is also supported by FCT under contracts UIDB/00618/2020, UIDP/00618/2020, CERN/FISPAR/0002/2017 and CERN/FIS-PAR/0014/2019.
This work has also been supported in part by the Swedish Research Council grant, contract number 2016-05996 and by the European Research Council (ERC) under the European Union's Horizon 2020 research and innovation programme (grant agreement No 668679).

\bibliographystyle{JHEP}
\bibliography{biblio}

\end{document}